\documentclass[twocolumn]{aastex631}

\usepackage{graphicx}
\usepackage{amsmath}
\usepackage{makecell}
\usepackage{natbib}
\usepackage{xcolor}
\usepackage{bm}
\usepackage{url}

\shorttitle{Nova Progenitors in M31}
\shortauthors{C.S. Abelson et al.}

\begin{document}

\title{The Progenitor Systems of Classical Novae in M31}

\author[0009-0007-2747-3691]{C.S. Abelson}
\affiliation{Department of Physics and Astronomy and Pittsburgh Particle Physics, Astrophysics and Cosmology Center (PITT PACC), University of Pittsburgh, 3941 O'Hara Street, Pittsburgh, PA 15260, USA}

\author[0000-0003-3494-343X]{Carles Badenes}
\affiliation{Department of Physics and Astronomy and Pittsburgh Particle Physics, Astrophysics and Cosmology Center (PITT PACC), University of Pittsburgh, 3941 O'Hara Street, Pittsburgh, PA 15260, USA}

\author[0000-0002-8400-3705]{Laura Chomiuk}
\affiliation{Center for Data Intensive and Time Domain Astronomy, Department of Physics and Astronomy, Michigan State University, East Lansing, Michigan 48824, USA}

\author[0000-0002-7502-0597]{Benjamin F. Williams}
\affiliation{Department of Astronomy, Box 351580, University of Washington, Seattle, WA 98195, USA}

\author[0000-0001-5228-6598]{Katelyn Breivik}
\affiliation{McWilliams Center for Cosmology, Department of Physics, Carnegie Mellon University, 5000 Forbes Avenue, Pittsburgh, PA 15213, USA}

\author[0000-0002-1296-6887]{L. Galbany}
\affiliation{Institute of Space Sciences (ICE-CSIC), Campus UAB, Carrer de Can Magrans, s/n, E-08193 Barcelona, Spain.}
\affiliation{Institut d'Estudis Espacials de Catalunya (IEEC), 08860 Castelldefels (Barcelona), Spain}

\author[0000-0002-4374-0661]{C. Jim\'enez-Palau}
\affiliation{Institute of Space Sciences (ICE-CSIC), Campus UAB, Carrer de Can Magrans, s/n, E-08193 Barcelona, Spain.}
\affiliation{Institut d'Estudis Espacials de Catalunya (IEEC), 08860 Castelldefels (Barcelona), Spain}

\begin{abstract}
We present the first characterization of the statistical relationship between a large sample of novae in M31 and their progenitor stellar populations in the form of a delay time distribution. To this end, we leverage the spatially resolved stellar age distribution of the M31 disk derived from deep \textit{HST} photometry by the Panchromatic Hubble Andromeda Treasury (PHAT) survey and a large catalog of novae in M31. Our delay time distribution has two statistically significant detections: one population of nova progenitors, ages between 2 and 3.2 Gyr, with an unnormalized rate of ($3.7^{+6.8}_{-3.5} \pm 2.1)  \cdot 10^{-9}$ events / $M_{\odot}$, and another of ages between 7.9 Gyr and the age of the Universe with ($4.8^{+1.0}_{-0.9} \pm 0.2) \cdot 10^{-9}$ events / $M_{\odot}$ (uncertainties are statistical and systematic, respectively). Together with the upper limits we derive at other time bins, these detections  are consistent with either a constant production efficiency or a higher production efficiency of novae at earlier delay times.
\end{abstract}


\section{Introduction} \label{sec: intro}
Novae are among the most common astrophysical transients arising from binary stellar evolution. However, no large-scale study of their progenitors has yet been conducted. A key concern in the study of binary systems is their effect on the formation and evolution of compact objects: in binaries with different initial masses, one star will always leave the main sequence first, giving rise to interactions between a living star and a stellar remnant. For example, when a star begins to transfer matter from its outer layers onto its degenerate companion -- whether by overfilling its Roche lobe or through strong stellar winds \citep{pod_moh_07} -- it creates the conditions for a thermonuclear explosion on the surface of the white dwarf. These explosions eject the envelope of accreted mass but are nonterminal for the remnant and the companion star. They often produce an optical transient, commonly known as a classical nova (\citealt{starrfield_72,gallagher_starrfield_78,prialnik_95,townsley_bildsten_04} -- see \citealt{chomiuk_21}, for a recent review). The details of the mass transfer process, the build-up to the thermonuclear runaway, and the consequences they both have on the subsequent evolution of the components of the binary system are complex. Inquests into this process, whether by stellar evolution codes \citep{denissenkov_12, denissenkov_14, paxton_15} or binary population synthesis (BPS) simulations \citep{chen_16, kemp_21}, are only possible by making considerable simplifying assumptions that significantly cloud our picture of the landscape of nova progenitors. 

Given their relative abundance compared to other astrophysical transients, novae provide a highly accessible probe into binary stellar evolution, enabling the use of a rich statistical toolset on larger samples to constrain the nature of their progenitor systems. A precise measurement of the evolutionary timescales and formation efficiencies of nova progenitors would provide observational constraints to help answer a number of open questions in binary stellar evolution. Among these questions are the influence of the initial conditions (since it is now well established that the fraction of close binary systems in the main sequence is a strong function of stellar properties like mass and metallicity; see \citealt{Moe2017,Badenes2018,moe_19,mazzola_20}), the impact of the orbital and stellar parameters on the stability of mass transfer \citep{Pavlovskii2017,temmink_23}, the role played by stellar winds in the mass transfer process \citep{Mohamed2007,webb_23}, and the many uncertainties involved in the onset, progression, and aftermath of common envelope episodes \citep{ivanova_11}. Many of these uncertainties are encoded explicitly or partially in BPS codes, which have been used to study novae by \cite{chen_16}, \cite{Chen2019}, \cite{kemp_21}, and \cite{Kemp2022}. A common thread in these theoretical studies has been a dearth of observational constraints derived from large, statistically significant samples of novae.

To address these issues, we present here the first measurement of the Delay Time Distribution (DTD) of novae in M31. The DTD is the occurrence rate of a class of objects as a function of time following a single brief burst of star formation. By characterizing the spatial correlation (or lack thereof) between the ages of stars and the objects of interest, we can recover the formation efficiency of that class of objects as a function of lookback time in field stellar populations -- another way of defining the DTD \citep{maoz_badenes_10,badenes_15,Sarbadhicary2021}. An observationally derived DTD can be used to test theoretical expectations, including the predictions from BPS models and their underlying assumptions.

The Andromeda galaxy (M31), the closest large galaxy to the Milky Way, is the ideal environment for a large-scale study of nova progenitors. At a distance of 752 $\pm$ 27 kpc \citep{riess_12}, the stellar populations in M31 can be resolved by \textit{HST} down to magnitude $\sim$27 in regions of low stellar density, which has allowed the Panchromatic Hubble Andromeda Treasury (PHAT) team \citep{dalcanton_12, williams_14,williams_17} to use precise multi-band photometry to produce spatially resolved maps of the stellar age distribution (SAD) in an area that encompasses roughly one third of the M31 disk. We have combined the PHAT data with the extensive historical catalog of novae in M31 from \citet{pietsch_07}, which contains over a thousand entries, to derive the DTD for novae in M31. This paper is structured as follows: Section \ref{sec: data} reviews the nova catalog and PHAT SAD map in detail, Section \ref{sec: methods} explains the theory behind the DTD and the process of recovering it, Section \ref{sec: results} presents our DTD in the context of the stellar isochrones used to generate the SAD map, and Section \ref{sec: conclusion} explores our results in more depth and compares them to previous literature.

\section{Data} \label{sec: data} 
Our work builds off decades of observations of M31; namely, a historical nova catalog covering the entire galaxy, originally created for comparison with an X-ray nova catalog \citep{pietsch_07}, but regularly updated in a publicly available website maintained by W. Pietsch\footnote{\url{https://www.mpe.mpg.de/~m31novae/opt/m31/}}. We combine these data with a spatially resolved ancient star formation history of M31 \citep{williams_17} produced from the PHAT survey \citep{dalcanton_12,williams_14}. This map provides the crucial link between the locations of novae and the age distribution of the stars surrounding them, enabling the recovery of a DTD.

\subsection{Nova catalog} \label{ssec: novae}
\begin{figure}
    \centering
    \includegraphics[width=\columnwidth]{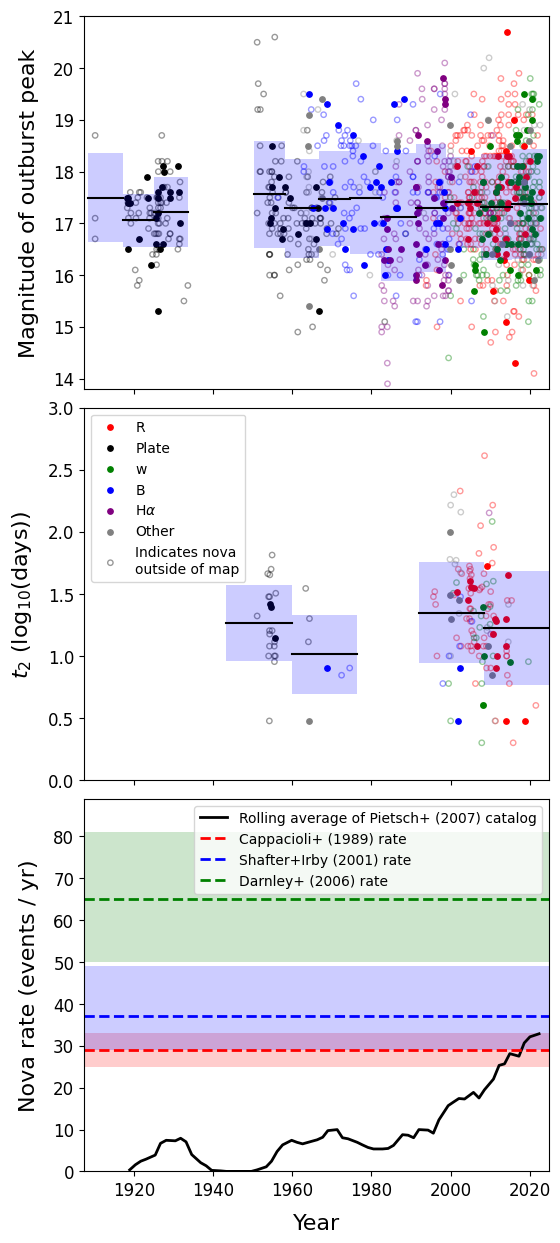}
    \caption{(a) Peak brightness in various filters vs. outburst date (used as a proxy for discovery date) for all novae in the Pietsch catalog. Novae that fall in the PHAT footprint are highlighted in pink. The average and standard deviation of the brightness in time bins of 3000 days are presented as the horizontal lines and shaded regions, respectively. (b) Time taken to decay from peak brightness to 2 magnitudes below peak brightness vs. outburst date. (c) 300-day rolling average of the observed nova rate as a function of outburst date, compared with estimations of the global nova rate in M31 from the literature(see text for details).}
    \label{fig: novae}
\end{figure}

The nova catalog goes as far back as the seminal survey of Andromeda led by Edwin Hubble \citep{hubble_29} and other surveys conducted using photographic plates \citep{mayall_31, payne_63, baade_63}. More recent additions come mainly from the Zwicky Transient Facility \citep{Bellm2019}, which scans the entire optical northern sky every 2 days. This high-cadence survey has a depth of 20.6 in the R band. The catalog is also populated by an array of observations from independent astronomers searching for transients and reporting their discoveries on databases such as the Central Bureau for Astronomical Telegrams and the Transient Name Server. By their nature, these disparate additions to the nova inventory are recorded in a variety of filters and limiting magnitudes. As of November 2022, the catalog contains 1219 individual entries, 263 of which fall in the footprint covered by the PHAT survey (see Section \ref{ssec: sad} below).

Because the nova catalog is heterogeneous and spans more than a century of observations, we must examine its contents critically. It is likely that some novae might have been missed in early studies or that faint and/or fast novae are underrepresented in surveys conducted before the advent of high-efficiency CCD detectors and high-cadence surveys, leading to biases and completeness issues. To evaluate these issues, we show the peak brightness and decay times of the novae in the Pietsch catalog as a function of discovery date in Figure \ref{fig: novae}, as well as a rolling average of the observed nova rate. The peak brightness of M31 novae ranges from magnitude 13.9 to 20.7, with an average and standard deviation of 17.3 $\pm$ 1.0. The decay times -- only reported for a subset of the sample -- span a much wider range between 2 and 410 days; taking the average and standard deviation in log space yields 1.28 $\pm$ 0.42 log(days).

To look for evidence of evolution in either the peak brightness or decay timescales as a function of discovery date, we divide the novae into temporal bins of 3000 days for the former and 6000 days for the latter (see Figure \ref{fig: novae}). Our analysis shows no statistical evidence for such evolution.

The measured nova rate has grown significantly in recent years, reflecting the increased efficiency of nova surveys. In panel c of Figure \ref{fig: novae}, we compare the rolling average of the nova rate to three estimates from the literature: those of \cite{cappacioli_89} and \cite{shafter_irby_01}, which focused on the historical record of observed novae in M31, and that of \cite{darnley_06}, who corrected for completeness using artificial nova tests and models of M31's surface brightness and internal extinction in the context of the POINT-AGAPE survey \citep{CalchiNovati2005}. The rolling average of the observed nova rate in the most recent epochs included in the \citeauthor{pietsch_07} nova catalog ($\sim$ 30 novae/yr) is consistent with the values derived by \cite{cappacioli_89} and \cite{shafter_irby_01}, but a factor two lower than the value in \cite{darnley_06}. This latter discrepancy is to be expected, as the rate reported in that study accounts for novae that would be unobservable and cannot be directly compared to observed rates. We conclude that while the \citeauthor{pietsch_07} nova catalog is certainly incomplete, it does not appear to have significant biases against any subclass of novae included in the sample, such as faint or fast-evolving novae.

Another property of the catalog entries relevant to our DTD analysis is the spatial accuracy of the nova positions. In theory, uncertainties on nova sky coordinates could propagate to uncertainties on nova counts in each spatial cell (see Section \ref{ssec: sad}). Unsurprisingly, novae discovered at earlier times tend to have larger uncertainties. The earliest members of the nova catalog were discovered on photographic plates, with their positions (and accompanying uncertainties) being reported as a Cartesian distance from some chosen center as opposed to sky coordinates \citep{shafter_15}. However, the median uncertainty of the entire catalog is half an arcsecond -- corresponding to 1.8 pc (projected) at the distance of M31, far smaller than the 300 $\times$ 1400 pc (deprojected) spatial cells of the SAD map -- and no nova has a coordinate uncertainty greater than 13'', or 47 pc (projected). We conclude that errors in the spatial location of M31 novae that could lead to ``cell hopping'' and propagate into uncertainties on the DTD are likely rare.

To remove contamination by recurrent novae (i.e., counting separate outbursts of the same system as different systems), we rely on \citet{shafter_15}, who screened the \citeauthor{pietsch_07} catalog looking for spatially coincident nova candidates in the original digital images or photographic plates. By removing all instances of recurrence reported by their analysis except for the most recent nova candidate, we ensure that each progenitor system gives rise to exactly one nova and every nova included in our DTD analysis is unique. Since this study was conducted, an additional 255 novae have been added to the \citeauthor{pietsch_07} catalog; to remove recurrent novae from these entries, we looked for those that showed spatial overlap within 1$\sigma$ of another nova. In total, our recurrence analysis removed 31 nova outbursts from the catalog, 10 of which fall within the PHAT footprint, leaving a total of 253 unique novae to be compared with the spatially resolved SAD. We also remedied an apparent typo in the outburst date of nova 2013-02a, recorded as Julian Date 3456332.98 (January 7th, 4751 CE), changing the leading digit from 3 to 2.

In order to report nova rates as well as abundances, we must account for the clear incompleteness of the catalog at early times. We attempt this correction in two ways; by deriving an effective survey length and by only analyzing a sample of recent novae, where the catalog is consistent with measured nova rates.

To calculate an effective survey length, we assume that the catalog is effectively complete in the final 3000-day time bin (see panel (a) of Figure \ref{fig: novae}), which corresponds to a rate of 30 novae / year -- in good agreement with other calculations of observed M31 nova rates \citep{cappacioli_89, shafter_irby_01}. We correct for incompleteness at earlier times with the following procedure: for each 3000-day time bin, we calculate the ratio of the nova count in that period and the count in the final period, multiply that ratio by the time span of the bin, and increment the total effective survey length by that diminished value rather than the full 3000 days. This calculation, which depends on the uncertain choice of a ``complete'' nova rate against which we compare our historical sample, yields an effective survey length of 38 years. We stress that this number is merely a best estimate -- as are the nova rates derived from it -- and heterogeneous historical catalogs such as this cannot be precisely corrected for completeness.

The most recent 6000 days of the \citeauthor{pietsch_07} catalog have an observed rolling nova rate consistent with previous studies (see Figure \ref{fig: novae}). By limiting our sample to these recent novae, we can provide a rough cross-check to our effective survey length completeness correction. However, this approach shrinks our nova sample considerably and limits our ability to recover the DTD. The results of both completeness analyses are presented in Section \ref{sec: conclusion}.

\subsection{PHAT stellar age distribution map} \label{ssec: sad}
\begin{figure}
    \centering
    \includegraphics[width=\columnwidth]{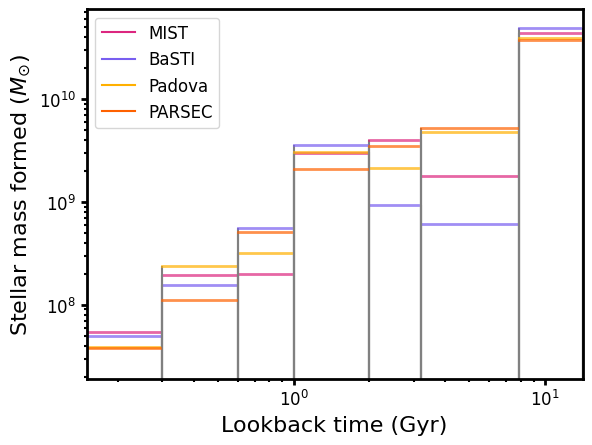}
    \caption{The global star formation history of M31, in units of mass formed, as measured by \citet{williams_17} and rebinned according to our temporal binning scheme. The different isochrone models yield slightly different measurements, with BaSTI emerging as an outlier in both the 5th and 6th bins. Due to their small relative size, uncertainties are omitted.}
    \label{fig: global sfh}
\end{figure}

The PHAT survey collected HST photometry for 117 million individual stars in the northern half of M31, measuring the stars simultaneously in 6 bands with coverage from the UV to the IR. This rich data set, hosted on MAST at \dataset[10.17909/T91S30]{http://dx.doi.org/10.17909/T91S30} \citep{MAST_data}, has resulted in dozens of publications that focus on different aspects of the stellar content of M31. \citet{williams_17} recovered the star formation history by dividing the PHAT data into 826 $83''\times83''$ (deprojected to 0.3$\times$1.4 kpc at the distance of M31) spatial cells, producing color-magnitude diagrams (CMDs) for each cell -- assuming a Kroupa initial mass function \citep{kroupa_02} -- that were then fit with the well-tested \texttt{MATCH} software package \citep{dolphin_02}. \texttt{MATCH} also contains tools to model both systematic \citep{dolphin_12} and random \citep{dolphin_13} uncertainties, which the authors implement when deriving their SAD map.

Certain features in a CMD are clear indicators of the presence of stars of a certain age; by comparing an observed CMD to a model composed of simple stellar populations of various ages, an SAD can be robustly recovered. \citet{williams_17} used MATCH 2.6 to calculate the best-fit SADs to CMDs drawn from several bands, only including stars brighter than the 50\% completeness limit and making use of the software's ability to incorporate complex extinction models. They draw from a high-resolution dust map of M31 \citep{dalcanton_15} derived from a two-component model of reddening-induced spread among RGB stars. This model provides a galactic map of median $A_V$, spread of the log-normal $A_V$ distribution, and fraction of stars being affected by dust; \citet{williams_17} then feeds these parameters to \texttt{MATCH}. They also account for foreground reddening by fitting an additional parameter $A_{VFG}$ independently for each spatial cell.

Each cell was fit using four different sets of isochrones: Padova \citep{marigo_08, girardi_10}, BaSTI \citep{pietrinferni_04, cassisi_05, pietrinferni_13}, PARSEC \citep{bressan_12}, and MIST \citep{choi_16}. This resulted in four maps of the SAD (or star formation history) in the PHAT footprint, one for each set of isochrones, each with 16 temporal bins spanning from the present day to the age of the Universe (see \citealt{williams_17} for details). The metallicities used to generate the isochrones were determined not by fitting a single $Z$ value to each spatial cell, but by modelling the enrichment history of the entire galaxy. Upon finding evidence for chemical enrichment at early times \citep{molla_97}, the authors make three radial divisions to M31 and fit an exponentially decreasing (in time) enrichment rate to each region. Their best-fit enrichment decay rates are independently consistent with regions of high star formation and the observed metallicity gradient in M31.

We assigned each of the 253 unique novae in the \citeauthor{pietsch_07} catalog that fall within the PHAT footprint to one of the 826 spatial cells in the \cite{williams_17} maps. The median PHAT cell has no novae, but 158 cells -- about 19\% -- have at least one, with the two most populated cells containing 9 distinct novae each. To find a good compromise between retaining enough temporal resolution in the DTD and maximizing the likelihood of detecting power in at least some of the combined bins, given the size of our sample, we combined the 16 time bins of the native PHAT SAD maps into 7 larger bins. We retained the youngest (0-300 Myr) and oldest (7.9-14.1 Gyr) temporal bins. The remaining native bins in the SAD maps, which vary in duration from 100 Myr to $>1$ Gyr, were merged as to be roughly equal in logarithmic time. This rebinning scheme improved the average ratio between the $M_{i,j}$, the stellar mass formed in spatial cell $i$ and time bin $j$, and $\sigma_{M_{i,j}}$ (the uncertainty on that quantity), while greatly reducing the number of spatial cells with $M_{i,j} = 0$. The SAD obtained with this temporal binning scheme, integrated over the entire PHAT footprint for each of the four isochrone sets is shown in Figure \ref{fig: global sfh}.

The four isochrone data sets used by \cite{williams_17}, and the corresponding SAD maps, offer a unique opportunity to evaluate the impact that systematic uncertainties associated with stellar evolution models have on the star formation histories recovered from resolved stellar populations. This topic is treacherous and uncertain \citep[see][for discussions]{Gallart2005,Conroy2009}, but it remains a fundamental limitation in attempts to draw statistical inferences about the association between resolved stellar populations and the products of stellar evolution, as we will show in Section \ref{sec: conclusion}.

\section{Methods} \label{sec: methods}

In recovering the DTD, this work follows the procedure laid out in \citet{maoz_badenes_10} and \citet{badenes_15}. Convolving the DTD ($\Psi(t)$, with units of events / yr / $M_{\odot}$) with the SAD ($M(t)$, with units $M_{\odot}$) yields an event rate over time $R(t)$ (units events / yr):
\begin{equation} \label{eqn: dtd convolution}
    R(t) = \int_0^t M(\tau) \Psi(t - \tau) d\tau \,.
\end{equation}
In practice, we only have access to the current event rate $R(t_0)$ and an SAD split into discrete time bins; however, the PHAT survey and M31 nova catalog gives us this information for 826 spatial cells. We can then recast the integral in Equation \ref{eqn: dtd convolution} discretely, as a dot product:
\begin{equation} \label{eqn: rate}
    R_i(t_0) = \sum_{j=0}^{t_0} M_{i,j} \Psi_j \,,
\end{equation}
where $R_i(t_0)$ is the modern-day nova rate in the i-th spatial cell, $M_{i,j}$ is the total stellar mass formed in the i-th cell and j-th temporal bin, and $\Psi_j$ is the DTD, or event rate per year per unit mass, in the j-th temporal bin (which is inherent to novae and therefore common across all spatial cells). The DTD recovery can now be treated as an inverse problem, where the values of the DTD in each temporal bin, $\Psi_j$, are free parameters that are varied to achieve the set of rates $R_i(t_0)$ that best fit the observed nova counts $n_i$ in all bins, simultaneously.

Typically, $R_i(t_0)$ would be calculated by dividing the number of novae in each spatial cell -- $n_i$ -- by the effective survey length \citep{maoz_badenes_10}. As noted in Section \ref{ssec: novae}, we estimate an effective survey length of 38 years for the \citeauthor{pietsch_07} catalog. However, due to the uncertainty in this estimate, we adopt $n_i$ as a proxy for $R_i(t_0)$ and report our DTD in units of events / $M_{\odot}$, nova count per unit of stellar mass, rather than nova rate per unit stellar mass (except where otherwise noted).

To explore this seven-dimensional parameter space and obtain Bayesian posteriors on each $\Psi_j$, we employ the dynamic nested sampling routine \texttt{dynesty} \citep{speagle_20, koposov_24}. In short, nested sampling \citep{skilling_04, skilling_06} is an efficient method of recovering posteriors $P(\theta | \text{Data}, \text{Model})$ on a set of parameters $\theta$ by simultaneously estimating the Bayesian evidence $P(\text{Data} | \text{Model})$ and the posterior. Dynamic nested sampling (DNS) \citep{higson_19} allocates the posterior samples adaptively throughout the sampling process, exploring high-likelihood regions of the parameter space more thoroughly. DNS has many attributes that make it an attractive alternative to methods like Markov Chain Monte Carlo, chief among them being its speed and the trivial independence of its samples.

In our application of DNS, we employ a log-likelihood function drawn from the modified chi-squared statistic derived in \citet{mighell_99}:
\begin{equation} \label{eqn: chi gamma}
    \chi_{\gamma}^2 = \sum_{i=0}^N \frac{(n_i + \textrm{min}(n_i, 1) - m_i)^2}{n_i + 1} \,,
\end{equation}
where $n_i$ is the observed number of novae in a spatial cell and $m_i$ is the expected number of novae according to our model. The latter number is calculated by convolving the model DTD with the observed SAD in the i-th spatial cell. We implement the Bayesian routine with the single bounding method of \citet{mukherjee_06} and the slice sampling method of \citet{neal_03}, \citet{handley_15a}, and \citet{handley_15b}.

The modified chi-squared statistic, $\chi_{\gamma}^2$, is designed to converge to the true mean value of Poisson-distributed data (even in the low-N regime), unlike the standard $\chi^2$. We employ this modified statistic rather than drawing a log-likelihood function directly from the Poisson distribution in order to directly incorporate variance in the expected number of novae $m_i$ deriving from the statistical variance in the SAD used to calculate it. Instead of propagating this uncertainty through bootstrap or hybrid Monte Carlo methods \citep{dolphin_13}, we exploit the fact that the denominator of the summed term in Equation \ref{eqn: chi gamma} corresponds to the variance of $n_i$ and add the variance on $m_i$, which is calculated through uncertainty propagation ($\textrm{Var}(m_i) = \sum_{j=0}^{t_0} \sigma_{M_{i,j}}^2 \Psi_j^2$).

We then minimize $\chi_{\gamma}^2$ by maximizing the log-likelihood of the corresponding Gaussian distribution:
\begin{multline} \label{eqn: log like}
    \textrm{ln(L)} = -\frac{1}{2} \sum_{i=0}^{N} \frac{(n_i + \textrm{min}(n_i, 1) - \sum_{j=0}^{t_0} \dot{M}_{i,j} \Psi_j)^2}{n_i + 1 + \sum_{j=0}^{t_0} \sigma_{M_{i,j}}^2 \Psi_j^2} + \\
    \textrm{ln} \left( 2 \pi \left( n_i + 1 + \sum_{j=0}^{t_0} \sigma_{M_{i,j}}^2 \Psi_j^2 \right) \right) \,,
\end{multline}
applying DNS to this equation and finding the parameters $\Psi_j$ that best fit our data. We note that this ``$\chi^2$-to-Gaussian'' approach is also taken by \citet{johnson_21} in their SED-fitting code \texttt{prospector}.

Any Bayesian routine requires the establishment of a prior distribution on the parameters being measured. We employ the same prior on all $\Psi_j$: flat between $0$ and $1.5 \cdot 10^{-6}$ events/$M_{\odot}$ and zero elsewhere. This upper limit is the rate necessary to produce the observed 1$\sigma$ upper limit on M31 nova rate \citep{darnley_06} in the case that the DTD is only nonzero in the single time bin with the least star formation.

For our final results, we calculated the edges of the 5\% highest probability density (HPD) region for each parameter and took the midpoint as the maximally likely rate. Similarly, our 1$\sigma$ uncertainty ranges on each parameter correspond to the 68.27\% HPD region, following \citet{badenes_15}. If the lower limit of this latter region coincides with the lower edge of the parameter space (technically, if the region falls within the smallest 100 parameter values explored by the \texttt{dynesty} fitting routine, with a typical DNS run involving upwards of $10^5$ such values), we report the rate to be statistically consistent with zero -- a non-detection -- and instead record a 2$\sigma$ upper limit.
  
\section{Results} \label{sec: results}
\begin{table*}[h]
    \centering
    \begin{tabular}{c|c c c c c|c}
    \vspace{0pt} & \multicolumn{5}{|c|}{Formation efficiency (events / $M_{\odot}$)} & \vspace{0pt} \\
    Time (Gyr) & MIST & Padova & PARSEC & BaSTI & \textbf{Combined (excl. BaSTI)} & ZAMS mass ($M_{\odot}$) \\
    \hline
    $10^{-3} - 0.3$ & $<4.3 \cdot 10^{-7}$ & $<6.6 \cdot 10^{-7}$ & $<7.1 \cdot 10^{-7}$ & $<5.2 \cdot 10^{-7}$ & \bm{$<6.2 \cdot 10^{-7}$} & $>3.2$ \\
    $0.3 - 0.6$ & $<2.5 \cdot 10^{-7}$ & $<2.5 \cdot 10^{-7}$ & $<4.1 \cdot 10^{-7}$ & $<3.6 \cdot 10^{-7}$ & \bm{$<3.2 \cdot 10^{-7}$} & $3.2 - 2.48$ \\
    $0.6 - 1$ & $<1.1 \cdot 10^{-7}$ & $<1.3 \cdot 10^{-7}$ & $4.7^{+2.4}_{-3.8} \cdot 10^{-8}$ & $6.5^{+3.0}_{-2.3} \cdot 10^{-8}$ & \bm{$<1.2 \cdot 10^{-7}$} & $2.48 - 2.06$ \\
    $1 - 2$ & $<1.9 \cdot 10^{-8}$ & $<2.1 \cdot 10^{-8}$ & $1.7^{+1.8}_{-1.0} \cdot 10^{-8}$ & $2.2^{+1.1}_{-1.4} \cdot 10^{-8}$ & \bm{$<4.1 \cdot 10^{-8}$} & $2.06 - 1.6$ \\
    $2 - 3.2$ & $5.1^{+6.8}_{-3.4} \cdot 10^{-9}$ & $<2.0 \cdot 10^{-8}$ & $3.7^{+6.2}_{-3.5} \cdot 10^{-9}$ & $<3.9 \cdot 10^{-8}$ & \bm{$(3.7^{+6.8}_{-3.5} \pm 2.1) \cdot 10^{-9}$} & $1.6 - 1.36$ \\
    $3.2 - 7.9$ & $<1.9 \cdot 10^{-8}$ & $<1.0 \cdot 10^{-8}$ & $<6.8 \cdot 10^{-9}$ & $<2.3 \cdot 10^{-8}$ & \bm{$<1.5 \cdot 10^{-8}$} & $1.36 - 1.06$ \\
    $7.9 - 14.1$ & $4.9^{+0.7}_{-0.8} \cdot 10^{-9}$ & $5.1^{+0.8}_{-0.8} \cdot 10^{-9}$ & $4.7^{+1.1}_{-1.2} \cdot 10^{-9}$ & $2.8^{+1.2}_{-1.1} \cdot 10^{-9}$ & \bm{$(4.8^{+1.0}_{-0.9} \pm 0.2) \cdot 10^{-9}$} & $1.06 - 0.90$ \\
    $\sum_i m_i$ & $240^{+100}_{-40}$ & $200^{+120}_{-30}$ & $250^{+90}_{-50}$ & $250^{+100}_{-70}$ & \bm{$210^{+150}_{-40} \pm 30$}
    \end{tabular}
    \caption{The maximum likelihood and 1$\sigma$ uncertainty ranges (for the combined DTD, the uncertainties given are statistical and systematic, respectively) of the posteriors for the DTD from each model (units of events/$M_{\odot}$). The ZAMS masses corresponding to the edges of each time bin are shown in the last column; these were calculated using MIST ``equivalent evolutionary point'' models with solar metallicity and $v/v_{crit}=0.4$. The final row displays the total expected number of novae in the footprint given by convolving the DTD and SAD given by the same model.}
    \label{tab: DTD}
\end{table*}

\begin{figure*}[h]
    \centering
    \includegraphics[width=\linewidth]{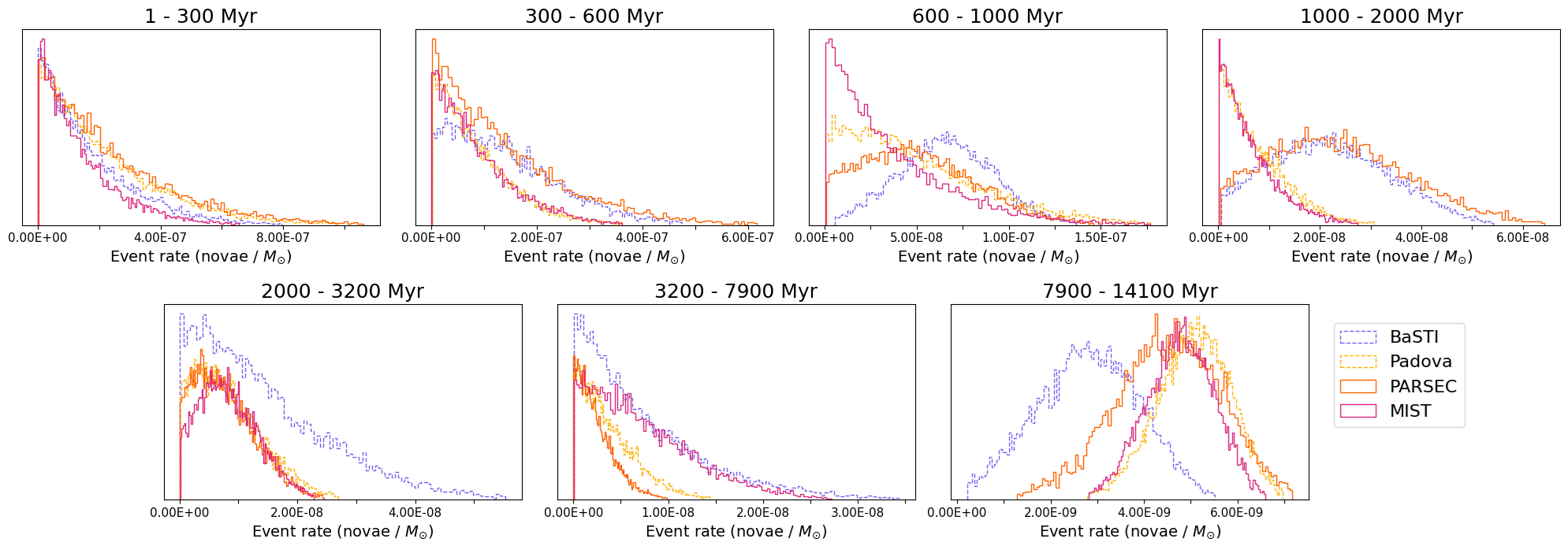}
    \caption{Posterior probability distributions on the DTD (in our chosen time binning scheme) for all four isochrone models. BaSTI is a notable outlier in its detection of a signal in the 600-1000 Myr bin, its lack of a detection in the 2-3.2 Gyr bin, and its disagreement on the value of the DTD in the 7.9-14.1 Gyr bin.}
    \label{fig: posteriors}
\end{figure*}

\begin{figure*}
    \centering
    \includegraphics[width=\textwidth]{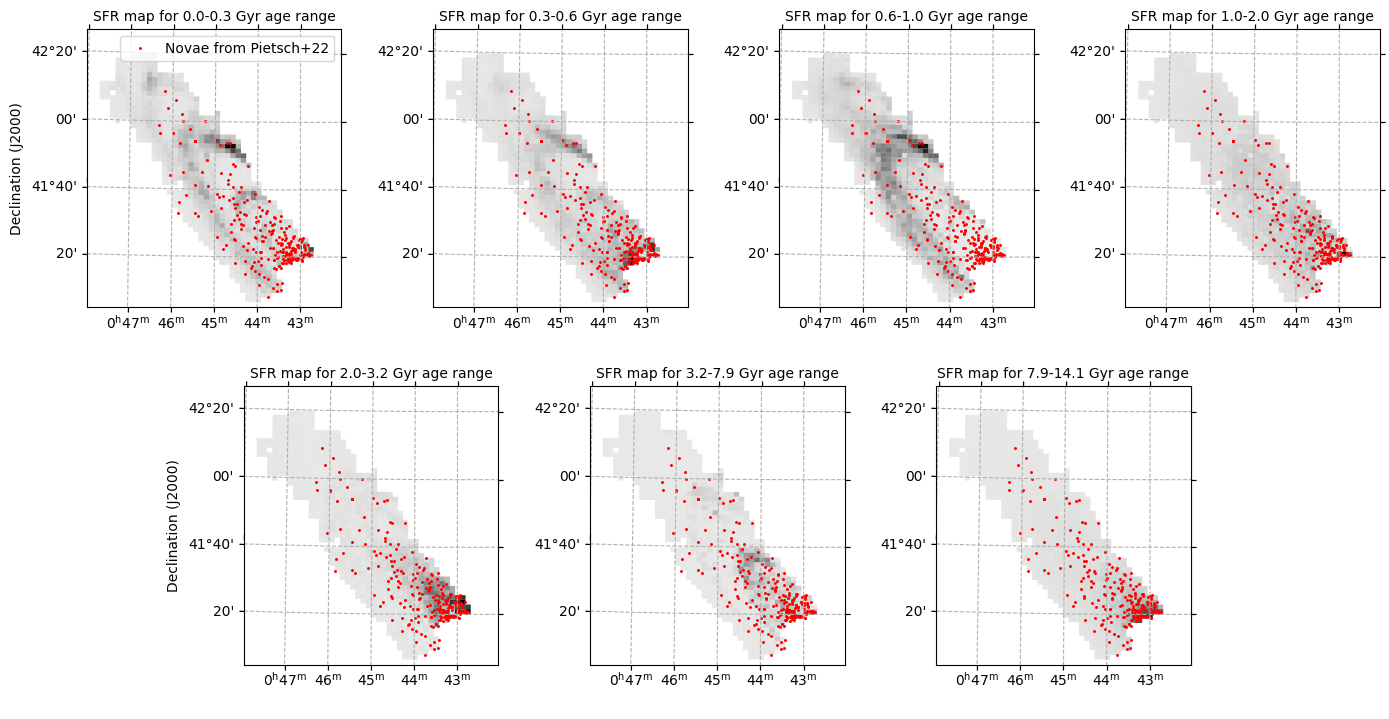}
    \caption{The PHAT SAD map of M31, calculated using MIST and binned according to our temporal scheme, with the Pietsch nova catalog overplotted. A darkly shaded spatial cell indicates higher SFR, with the shading normalized separately for each time bin.}
    \label{fig: sad map}
\end{figure*}

In Figure \ref{fig: posteriors}, we present the posteriors on event rate per unit stellar mass in each time bin, computed separately for the four available isochrone models. The DTD is a mathematical representation of the spatial correlation between events and stellar populations, as illustrated in Figure \ref{fig: sad map}. In this figure, produced using the MIST isochrone set, there is a clear correlation between nova positions and stellar mass of ages 2-3.2 Gyr and 7.9-14.1, which indicates the possibility of a physical link between those stars and that progenitor population of novae. We expect to see -- and indeed do see -- significant detections in the DTD corresponding to these temporal bins. We also note that our DTD is dissimilar to a DTD generated from random sky positions, showing that these detections are specific to novae.

The most prominent feature of these posteriors is the robust recovery of a population of nova progenitors in the oldest time bin, with ages between 7.9 Gyr and the age of the Universe, though BaSTI yields a detection that is $\sim 40\%$ lower than the average of the other three. This divergence is unsurprising, as in \citet{williams_14}, BaSTI was found to be most discrepant from the other models at very old ages. According to MIST stellar evolution models, these lifetimes correspond to zero-age main sequence (ZAMS) masses of $1.06 - 0.90$ $M_{\odot}$. With the exception of BaSTI, the other three isochrone sets (MIST, Padova, and PARSEC) also yield a detection in the 2-3.2 Gyr time bin. BaSTI and PARSEC also detect a signal in both the 0.6-1 Gyr and 1-2 Gyr time bins, but MIST and Padova do not. Disagreements at these young ages are not unexpected; the models become less accurate as the main sequence turnoff point falls below the photometric depth of the PHAT survey. For all other time bins, the maximally likely event rate was either less than the standard deviation of the posterior (which we treat as a nondetection) or essentially zero.

There is broad agreement between the isochrone sets, and the results are consistent with our expectation of higher formation efficiencies at earlier delay times from BPS models (discussed further in Section \ref{sec: conclusion}). We note that the exact values, not just the general features, of the DTDs from each isochrone set are broadly consistent when \texttt{dynesty} is run multiple times on the same SAD map. For our final DTD measurement, we take statistical and systematic uncertainties into account by averaging only the results of the PARSEC, Padova, and MIST posteriors and combining the standard deviations of the three posteriors in quadrature, then presenting these uncertainties alongside the standard deviations between the three models (see Table \ref{tab: DTD}). Having excluded BaSTI from the final DTD, we can see that the statistical uncertainties are larger than the systematic differences between isochrone sets, although the range of variation is significant: the ratio of statistical to systematic uncertainty is $\sim$2:1 in the 2.0-3.2 Gyr bin, but $\sim$9:1 in the 7.9-14.1 Gyr bin. The final row of Table \ref{tab: DTD} shows that, as anticipated, the expected number of novae in all spatial cells derived from our recovered DTD, $\sum_i m_i$, falls within 1$\sigma$ of the observed number of unique novae in the PHAT footprint $\sum_i n_i$, which is 253.

The most notable disagreements between the DTDs derived from the four isochrone sets, in the time bins where they do differ, can be traced back to disagreements on the location of star formation in specific time bins. An illustrative example of this is the 600-1000 Myr bin shown in Figure \ref{fig: posteriors}. The signal grows more statistically significant in the order MIST, Padova, PARSEC, BaSTI -- corresponding to the sequence of increasing stellar mass formed in that same time bin (see Figure \ref{fig: global sfh}).

\begin{figure*}[h]
    \centering
    \includegraphics[width=\linewidth]{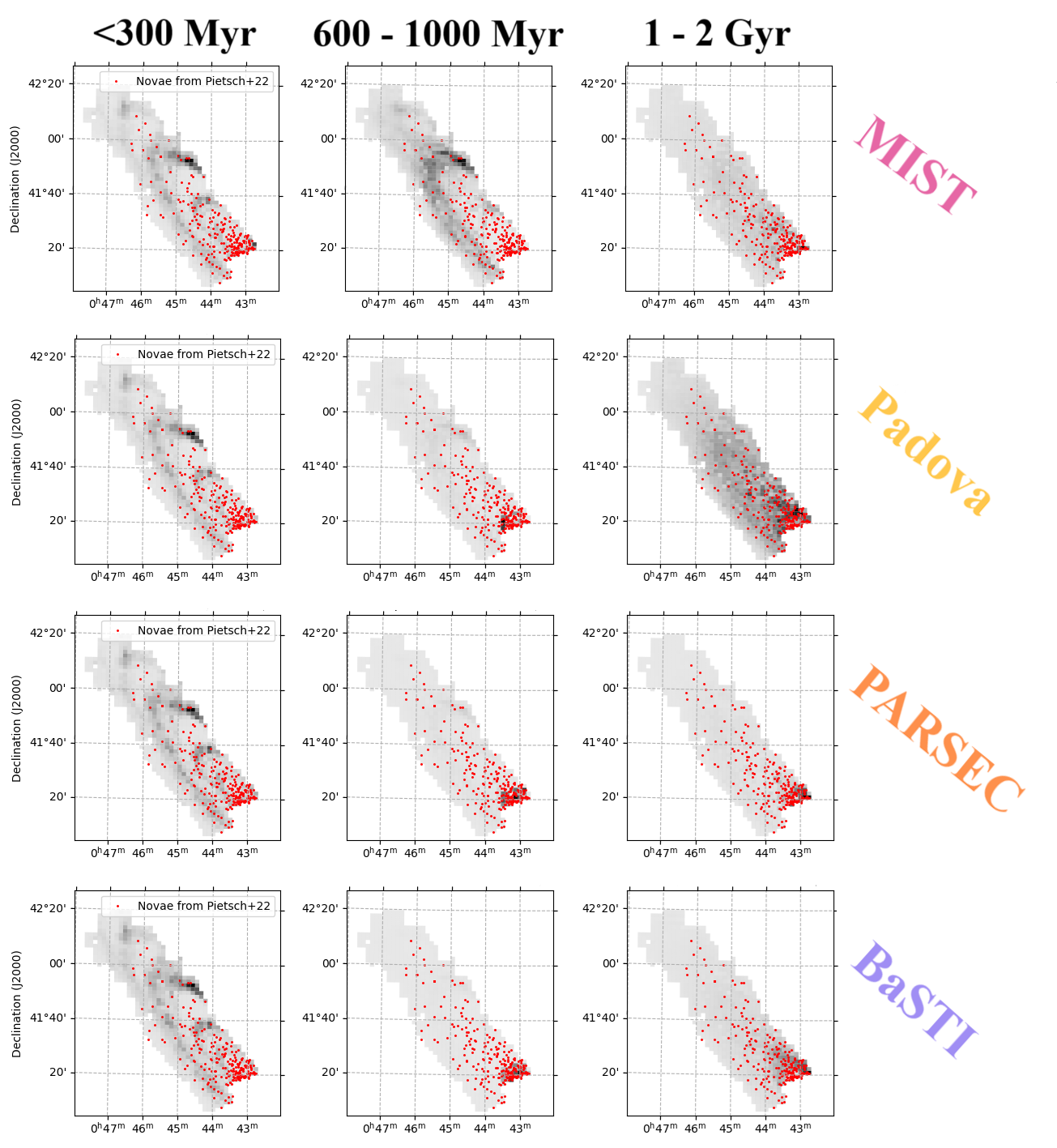}
    \caption{Comparisons between the SAD map derived from the four isochrone models for select time bins. The gray scale is normalized independently for each plot.}
    \label{fig: sad comparison}
\end{figure*}

Figure \ref{fig: sad comparison} further clarifies this disagreement. The concentration of star formation in the 10 and 20 kpc rings, which is present in all four models in the first time bin (0 -300 Myr), yields a consistent nondetection in the corresponding bin of Figure \ref{fig: posteriors}, given that the spatial distribution of novae shows no enhancement at these locations. However, the varying amount of star formation in the 20 kpc ring recovered by each isochrone set in the 600-1000 Myr range leads to the noted discrepancies in the recovered DTD. In the 1-2 Gyr time bin, the MIST and Padova models detect substantial star formation in the outer regions of the disk, whereas PARSEC and BaSTI only detect it close to the bulge -- colocated with the vast majority of novae. For this reason, the latter two models report a statistically significant nova rate in the corresponding time bin of the DTD, and the former two do not. These discrepancies between the DTDs derived using the different isochrone sets stress the model-dependent nature of DTD analyses and the importance of taking into account systemic biases and uncertainties.

\begin{figure*}[ht]
    \centering
    \includegraphics[width=\textwidth]{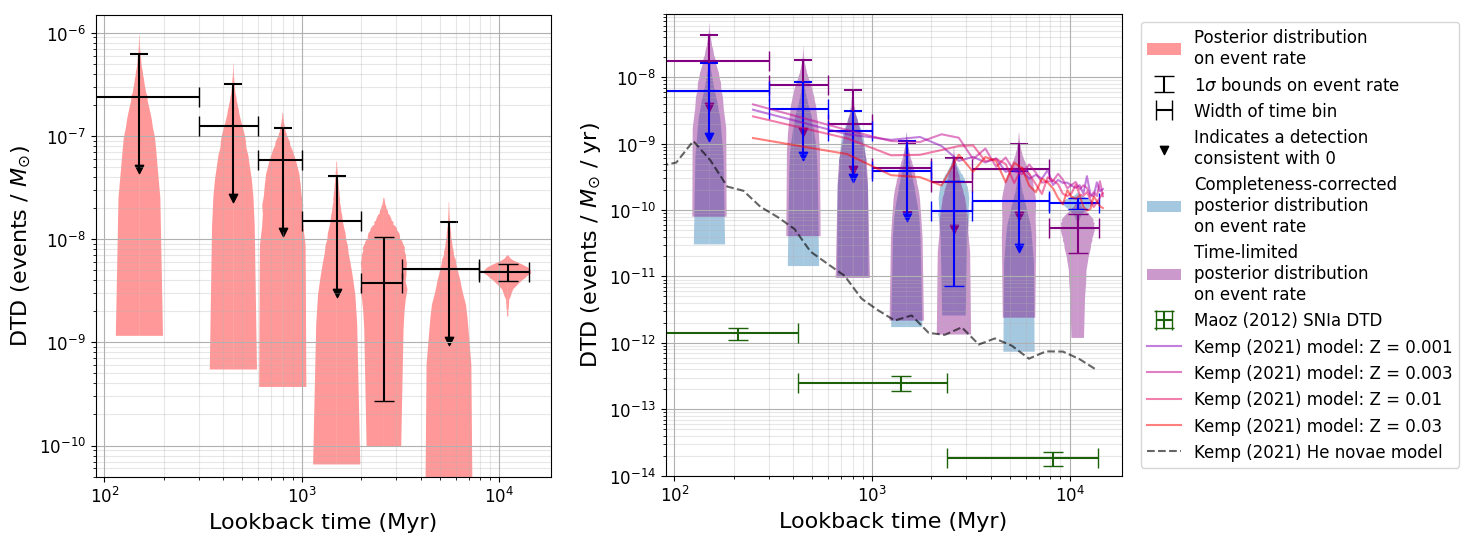}
    \caption{(a) Violin plot of our posterior distributions for the DTD rate in each time bin. (b) Violin plot of our completeness-corrected DTDs plotted against theoretical DTDs from \citet{kemp_21} and an observed SNIa DTD from \citet{maoz_12}. The DTD in blue is derived using the effective survey length; the DTD in purple is derived using the time-limited subset of the nova sample.}
    \label{fig: comparison}
\end{figure*}

In Figure \ref{fig: comparison}, we present a violin plot of our final DTD (in events / M$_{\odot}$) and two completeness-corrected DTDs; one obtained by dividing our results by the effective survey length of 38 years obtained in Section~\ref{ssec: novae}), and one obtained by limiting our sample to recent novae. These two DTDs are statistically consistent with each other except in the final time bin, where the 1$\sigma$ bounds come close to overlapping, but do not. We compare these completeness-corrected DTDs with a set of theoretical DTDs from \citeauthor{kemp_21}, as well as an observationally determined DTD for Type Ia SNe from \cite{maoz_12}, which we discuss in the following section.

\section{Discussion and Conclusions} \label{sec: conclusion}

We have presented the first statistical inference of the formation efficiency of novae across stellar populations of different ages (i.e., a DTD for novae) in M31. Our results are derived using SAD maps in M31 from deep HST photometry produced by the PHAT collaboration \citep{williams_17}, and the catalog of M31 novae compiled by W. Pietsch, which spans more than 100 years of observations. Taking statistical and systematic uncertainties into account, our DTD has robust detections in two time bins: stars between 2 and 3.2 Gyr of age and between 7.9 Gyr and the age of the Universe. While our ability to recover signal in certain time bins may be due to the particular nature of M31's star formation history and nova catalog, the DTD itself is intrinsic to the event of interest and, critically, does not depend on the data used to recover it.

The fact that M31's global SAD is dominated by older stars (see Figure \ref{fig: global sfh}) naturally makes the recovery of a DTD signal at young ages more difficult, particularly with a sample of modest size. Additionally, the excess of stellar mass in the 2-3.2 Gyr age range facilitates a detection at these delay times; one possible explanation for this signal is a burst in nova progenitor formation driven by the 2-3 Gyr old galactic merger \citep{hammer_18}. In this context, the question of which stellar populations generate most of the novae in M31 is distinct from the question of which stellar populations are most efficient at generating novae in general.

Taken together, our detections and upper limits are consistent with a decaying rate of nova formation with delay time; though, strictly speaking, we cannot rule out a uniform formation rate. The maximally likely rate of formation efficiency we recover is, perhaps surprisingly, higher in the final time bin than in the 2-3 Gyr bin, possibly implying a departure from monotonic decay of nova formation. In fact, there is no theoretical expectation that the DTD of novae should be exactly monotonic - the theoretical DTD calculated in \citet{kemp_21} exhibits several local maxima. However, given the wide range of rates falling within our 1$\sigma$ uncertainty intervals, we caution against over-interpretation of this result.

A decaying formation rate is borne out by simple arguments in stellar evolution. Technically, a binary system can produce novae as soon as the more massive partner evolves into a white dwarf. As delay time increases, the zero-age main sequence masses of both the donor and the degenerate accretor decrease on average, leading to a decrease in formation efficiency of novae \citep{ritter_91, kolb_95, chen_16} and an increase in recurrence time \citep{yaron_05}, as less mass is available to transfer to the accretor. Although the PHAT SAD map lacks sufficient resolution at short delay times to probe the ``turn-on'' phase -- the prediction that the formation efficiency of novae should be zero before the formation of the first WDs at delays of $\sim$40 Myr --, it does probe, and successfully recover, the broad expectation of a decrease in formation efficiency at longer delays.

Our rough completeness corrections allow us to estimate nova rates (per unit stellar mass) as well as nova counts; these rates are consistent with the shape and height of the \citeauthor{kemp_21} DTD (see Figure \ref{fig: comparison}), with the effective survey length correction yielding a slightly better match than the time-limited sample. To be clear, no adjustments have been made to our data to bring them into alignment with the theoretical rates. This agreement is simply a product of the DTD recovery process outlined in Section \ref{sec: methods} and the completeness corrections presented in Section \ref{ssec: novae}. In the same figure, we also make a comparison against the \citeauthor{kemp_21} DTD of helium novae -- rare events caused either by the deposition of helium, rather than hydrogen, onto a white dwarf, or the fusion of hydrogen into helium on the surface of the dwarf before a nova detonation. As expected, He nova rates are so low as to be inconsistent with our completeness-corrected DTD; this large discrepancy implies that a minute fraction of novae in M31 are He novae.

Previous BPS studies of novae, such as \citet{chen_16} and \citet{kemp_21}, provide theoretical support for more efficient nova production at earlier times. The claim made by \citeauthor{chen_16}, that less massive binaries will generate fewer novae, is borne out by their nova rate over time for two model SFHs. \citeauthor{kemp_21} go into much greater depth with regard to the progenitor population, providing detailed descriptions of common evolutionary pathways and their implications for nova rates. They present a (downward-sloping) DTD of their own and split it up by the evolutionary phase of the donor star. Following from Figure 12 from \citet{kemp_21}, the progenitor accretor population of the $2-3.2$ Gyr time bin should be dominated by O/Ne WDs. The progenitor donor population is a more eclectic mix, with low-mass main sequence and first giant branch stars best represented. The accretor population in the $7.9-14.1$ Gyr bin should be an approximately even mix of O/Ne and C/O WDs, with first ascent giant branch stars dominating the donor population.

Studies of novae, both in our galaxy and in Andromeda, have been plagued by dust-driven spatial incompleteness, the extent of which has been a subject of ongoing debate. M31's bulge-dominated nova population stands in contrast to other galaxies \citep{ciardullo_89} and the previously discussed expectation that younger stellar populations should be more efficient progenitors of novae. Were our catalog to be biased against novae in the disk, any correlation between young stellar populations and novae would be confounded, artificially driving down the DTD at early delay times. \citet{Hatano1997} used a simple model of dust in M31 to argue that the observed bulge-to-disk nova ratio is a consequence of dust extinction in the disk, and the true fraction of novae residing in the bulge is of order 25\%. However, their conclusion is deeply model-dependent; even their limited explorations of changes to this model are consistent with a bulge-dominated nova population. Notably, a map of extinction in M31 created from the PHAT survey \citep{dalcanton_15} does not resemble the \citeauthor{Hatano1997} model.

\citet{shafter_irby_01} use the spatial distribution of planetary nebulae in M31, which should be slightly more sensitive to extinction than novae, as a tracer of dust. They find this population to be less centrally concentrated than novae, concluding that the bulge dominance of the nova population in M31 is genuine and not an artifact of the dusty disk. There is no doubt that the \citeauthor{pietsch_07} catalog misses some novae due to dust. However, the result from \citeauthor{shafter_irby_01} and the fact that our results are consistent with higher formation efficiency among younger stellar populations make the existence of a large missing population of disk novae that would introduce severe spatial incompleteness unlikely.

Recent literature suggests a population of ``faint, fast'' novae \citep{shara_17} that would evade surveys with cadences much larger than an hour. This prediction, also supported by the theoretical models of \citet{yaron_05}, has not been borne out in newer surveys. Between 2018 and 2019, the Zwicky Transient Facility surveyed fields in the galactic plane with a typical cadence of 40 seconds to search for short-period astrophysical variables \citep{kupfer_21}; no such novae have been reported. \citeauthor{shara_17} predicts that, at the distance of M31, these objects would have magnitude 17-18 and decay times $t_2$ of 5 hours or less. Our nova catalog certainly probes these magnitudes, but appears not to be sensitive to such short delay times (see Figure \ref{fig: novae}). Therefore, our DTD does not contain information about these transients, which may not even comprise a significant fraction of novae.

Ultimately, one can only ever produce a DTD for observed events. The kinds of light-curve and extinction completeness corrections undertaken in studies of absolute nova rates can estimate the total number of novae left unobserved, but naturally cannot estimate the location of each unobserved nova -- the data that would be required to recover a DTD.

We conclude with a brief discussion of the comparison between our nova DTD and the DTD for Type Ia SNe recovered by \cite{maoz_12}, which is interesting as some subclasses of novae have been proposed as potential SN Ia progenitors \citep{Maoz2014}. Our completeness-corrected nova DTD is about four orders of magnitude higher than the SN Ia DTD, and noticeably shallower at long delay times. This implies a strong ($\lesssim0.1\%$) upper limit on the fraction of nova-producing systems that go on to explode as SN Ia in star-forming galaxies like M31.

Such a discrepancy presents a problem for the single-degenerate theory of Type Ia progenitors, which posits that a significant fraction of SNe Ia arise from gradual mass transfer between a living star and a remnant -- the same kind of system that produces novae. If such a small fraction of these systems end their lives as type Ia supernovae (right panel of Figure \ref{fig: comparison}), there may not be enough single-degenerate systems in star-forming galaxies to explain the observed Ia rate \citep{Maoz2014}.

Our work reinforces the previously demonstrated power of DTDs for constraining the progenitor populations of the products of binary stellar evolution. The continuation of high-cadence surveys of nearby, well-studied galaxies such as M31 would provide better constraints on the nova DTD at all lookback times and reduce our reliance on historical nova catalogs of dubious completeness. Resolved stellar populations of those galaxies provide opportunities for the recovery of DTDs for other astrophysical transients. The upcoming extension of the PHAT SAD map to the southern region of M31 \citep{chen_25} promises a near-trivial doubling of the nova sample size; once published, this data will represent an exciting opportunity to improve the resolution of the nova DTD.

SAD maps similar to that of \citet{williams_17} exist for other galaxies in the Local Group (e.g., \citealt{Lazzarini2022} for M33, \citealt{Mazzi2021,Harris2009} for the LMC, \citealt{Rubele2018,Harris2004} for the SMC), with matching, high-quality catalogs of astrophysical transients and other products of stellar evolution. This research enables further application of the methods described here, which could be leveraged to constrain BPS models, deepening our understanding of key stages of binary stellar evolution and improving the predictions for the rates of rare events such as SNe Ia and black hole mergers.

The data and code underlying this work will be shared upon a reasonable request made to the lead author.

We acknowledge fruitful discussions with Dan Maoz, Peter Nugent, and Robin Ciardullo. We are especially grateful to Alex Kemp, who kindly provided us with a digital form of his nova DTD. L. Galbany acknowledges financial support from AGAUR, CSIC, MCIN and AEI 10.13039/501100011033 under projects PID2023-151307NB-I00, PIE 20215AT016, CEX2020-001058-M, and 2021-SGR-01270. This research was supported in part by the University of Pittsburgh Center for Research Computing, RRID:SCR\_022735, through the resources provided. Specifically, this work used the HTC cluster, which is supported by NIH award number S10OD028483. We acknowledge financial support from NSF grant AST-2307865, and HST grants HST-GO-16741.002-A, HST-GO-16287.002-A, and HST-GO-17179.002-A. The following Python packages were a significant part of this research: Numpy \citep{numpy}, Matplotlib \citep{matplotlib}, and Astropy \citep{astropy_13, astropy_18, astropy_22}.

\bibliography{bib}
\bibliographystyle{aasjournal}

\end{document}